\documentclass [onecolumn,10pt,epsf]{article}

\usepackage[dvips]{graphics}
\usepackage{epsfig}
\usepackage{amsmath}

\makeatletter

\def\@normalsize{\@setsize\normalsize{10pt}\xpt\@xpt
\abovedisplayskip 10pt plus2pt minus5pt\belowdisplayskip
\abovedisplayskip \abovedisplayshortskip \z@
plus3pt\belowdisplayshortskip 6pt plus3pt
minus3pt\let\@listi\@listI}

\def\subsize{\@setsize\subsize{12pt}\xipt\@xipt}

\def\section{\@startsection {section}{1}{\z@}{1.0ex plus 1ex minus .2ex}{.2ex plus .2ex}{\large\bf}}

\def\subsection{\@startsection {subsection}{2}{\z@}{.2ex plus 1ex} {.2ex plus .2ex}{\subsize\bf}}
\makeatother

\begin{document}

\title{Spin-Orbit Interaction in particles' motion \\ as a model of quantum computation
}
\author{Gerasimos G.Rigatos \\ \\
\small Unit of Industrial Automation \\
\small Industrial Systems Institute \\
\small 26504, Rion Patras, Greece \\
\small e-mail: grigat@isi.gr}

\date{}

\maketitle

\noindent {\textbf{Abstract}:} The paper studies spin-orbit interaction  (i.e. the effect the spin has on the particle's trajectory in a magnetic field) as a model of quantum computation. The two-level spin quantum system is examined using the stochastic mechanics formulation. The control of the entangled spin state is considered as a problem of control of the mean moment of a particles ensemble along a reference trajectory. It is shown that such a control can be succeeded by applying an open-loop control scheme. \\

\section{Introduction\\}

\noindent The spin states, i.e. the eigenstates associated with the measurement of a particle's magnetic moment along the $z$-axis are a fundamental element of highly entangled states. For example, in quantum computing all the unitary entangle operations on many spins can be implemented by compositions of those on the two spins (universality of quantum circuits). Therefore, the stabilization of the two-spin system is of great importance for designing a quantum entangler machine. There are two types of control for chaotic and quantum systems, i.e. open-loop control where the controls are predetermined at the start of the experiment and closed-loop (or feedback control) where the control can be chosen through-out the experiment [\ref{FraEva05}].  Previous work on quantum open-loop control includes flatness-based control on a single qubit gate [\ref{SilRou08}]. In this paper the classical Schr\"{odinger's} equation is used and quaternion notation for the spin's Hamiltonian is applied. Then a flatness-based controller is derived for steering the particle's transition between the spin's eigenstates. On the other hand, quantum feedback control was developed as a quantum analogy to the classical theories of nonlinear (Stratonovich) filtering [\ref{Bel83}]. This approach is based on an analogous of the separation principle which holds in classical stochastic control [\ref{Ben91}]. Quantum feedback control is actually observer-based control. First quantum filtering is used to obtain an estimation of the stochastic quantum variable and then a feedback controller is designed based on the output of the quantum filter. The quantum filter describes a classical stochastic process [\ref{YamTsuHar07}],[\ref{DasRoy06}],[\ref{MirRou04}].\\

\noindent The present work studies spin-orbit interaction as a model of quantum computation (i.e. the effect the spin has on the particle's trajectory in a magnetic field) and examines control of the two-level spin quantum system using the stochastic mechanics approach. In stochastic mechanics each particle follows a continuous path  which is random but has a well-defined probability distribution [\ref{kleb05}],[\ref{ComMey06}]. Stochastic mechanics has been recognized as a self-consistent formulation of quantum mechanics in the framework of stochastic processes and is established as part of stochastic control theory [\ref{Far82}],[\ref{Far06}],[\ref{MarSieIllu97}],[\ref{Fli07}]. The particles' kinematic model is associated to the model of a quantum oscillator, taking the particle to be a 3-DOF variable (position in a cartesian coordinates frame) [\ref{Rig08}],[\ref{RiTza02}]. When the effect of the spin in the particle's motion is also considered, and the particle's motion takes place under an external  magnetic field (as described in the Stern-Gerlach experiment) then the deflection of the particles' trajectory should be also taken into account  [\ref{Far82}],[\ref{TanDiuLal98}].  The component of the spin along the field becomes a discrete random variable which is correlated with the average velocity [\ref{Far82}].\\

\noindent The concept adopted in this paper is that starting from Schr\"{o}dinger's equation and passing through the Fokker-Planck equation and the Ornstein-Uhlenbeck diffusion one finally arrives at Langevin's stochastic differential equation. A form of Langevin's equation is also used to describe the variations between the spin-eigenstates (discrete energy levels of the spin system) [\ref{Far82}],[\ref{Far06}]. This enables to consider the control of the entangled spin state as a problem of motion control of the mean of a particles ensemble along a reference trajectory. It is shown that using an appropriate open-loop control scheme (which is based on flatness-based control theory and which can be realized by a magnetic field) motion control of the particles mean along the desirable trajectory is possible. This in turn implies that the quantum state represented by the particles moment can be also controlled.

\section{The spin as a two-level quantum system\\}

\subsection{Description of a particle in spin coordinates\\}

\noindent The basic equation of quantum mechanics is
\textit{Schr\"{o}dinger's equation}, i.e.

\begin{equation} \label{Schrodingers_equation}
i{{{\partial}{\psi}} \over {{\partial}t}}=H{\psi(x,t)}
\end{equation}

\noindent where $|\psi(x,t)|^2$ is the probability density
function of finding the particle at position $x$ at time instant
$t$, and $H$ is the system's Hamiltonian, i.e. the sum of its
kinetic and potential energy, which is given by $H={p^2}/{2m}+V$,
with $p$ being the momentum of the particle, $m$ the mass and $V$
an external potential. The solution of Eq.
(\ref{Schrodingers_equation}) is given by $\psi(x,t)=e^{-iHt}{\psi(x,0)}$
\ref{TanDiuLal98}.\\

\noindent However, cartesian coordinates are not sufficient to describe the particle's behavior
in a magnetic field and thus the spin variable taking values in SU(2) has been introduced. In that case the solution
$\psi$ of Schr\"{o}dinger's equation can be represented in the basis $|r,\epsilon>$ where $r$ is the position vector
and $\epsilon$ is the spin's value which belongs in $\{-{1 \over 2},{1 \over 2}\}$ (fermion). Thus vector $\psi$ which appears in Schr\"{o}dinger's equation can be decomposed in the vector space $|r,\epsilon>$ according to $
|\psi>=\sum_{\epsilon}{\int}{{d^3}r}|r,\epsilon>,<r,\epsilon|\psi>$. The projection of $|\psi>$ in the coordinates system $r,\epsilon$ is denoted as $<r,\epsilon|\psi>=\psi_{\epsilon}(r)$. Equivalently one has $\psi_{+}(r)=<r,+|\psi>$ and $\psi_{-}(r)=<r,-|\psi>$. Thus one can write $\psi(r)=[\psi_{+}(r), \psi_{-}(r)]^T$.\\

\subsection{Motion of the particle in a magnetic field\\} \label{subsection : particle_motion_magnetic_field}

\noindent The interaction between the particle's spin and a magnetic field and the effect this has on particles' motion is described in the Stern-Gerlach experiment.  It can be observed that due to the gradient of the magnetic field and the particle's spin a deviation of the particle's trajectory takes places. It is assumed that the intensity of the magnetic field $B_z$ is positive while its gradient ${{\partial{B_z}} \over {{\partial}z}}$ is negative.\\

\noindent Since the particles are taken to be neutral they are not subject to Laplace forces $F=q{\cdot}v{\times}B$. The particles however have magnetic moment $M$ which is associated to potential energy given by $W=-M{\cdot}B$. There is also the kinetic moment $\Gamma$ which is related to the magnetic moment $M$ according to the relation $M={\gamma}{\cdot}{\Gamma}$, where $\gamma$ is the gyromagnetic ratio. The force that is applied to the particle is the gradient of the particle's potential energy $F={\nabla}(M{\cdot}B)$. Between the kinetic moment $\Gamma$ and the magnetic moment $M$ the following relation also holds
${{{\partial}\Gamma} \over {{\partial}t}}=M{\times}B{\Rightarrow}{{{\partial}\Gamma} \over {{\partial}t}}={\gamma}{\Gamma}{\times}B$. The particle behaves like a gyroscope. The term ${{{\partial}\Gamma} \over {{\partial}t}}$ is perpendicular to $\Gamma$ while the kinetic moment rotates round the magnetic field $B_z$. It can be seen that the kinetic moment $\Gamma$ is proportional and collinear to the magnetic moment $M$. According to $F={\nabla}(M{\cdot}B)$ the magnetic field $B_z$ forces the magnetic moment $M$ to rotate round it, with constant angular velocity. To calculate force $F$ the gradient of the particle's potential energy is found

\begin{equation} \label{gradient_potential_energy}
\begin{tabular}{c}
$W=-M{\cdot}B=-{M_x}{B_x}-{M_y}{B_y}-{M_z}{B_z}$
\end{tabular}
\end{equation}

\noindent It can be assumed that $M_z$ and $M_x=M_y=0$. The latter ir true because the frequency of rotation of $M$ is very high, and it is not possible for $M_x$ and $M_y$ to affect the particle's motion in any other way than their average value, which is $0$. Then it holds

\begin{equation} \label{force_of_magnetic_field}
F=\nabla(M{\cdot}B)={M_z}{\nabla}{B_z}
\end{equation}

\begin{figure}[htb]
\begin{center}
\rotatebox{-90}{\epsfig{file=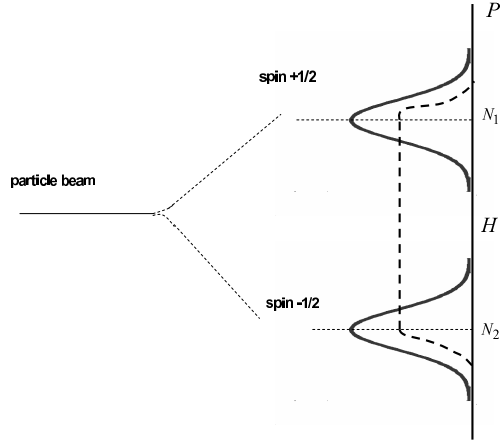, width=70mm,
height=90mm}} \caption{The two distributions denote that the magnetic moment of the particle can take only two distinct values
(spin up or spin down)}
\label{fig : spin_distributions}
\end{center}
\end{figure}

\noindent It also holds that ${{{\partial}{B_z}} \over {{\partial}x}}=0$ and ${{{\partial}{B_z}} \over {{\partial}y}}=0$ because the magnetic field is assumed to be independent of $x$ and $y$. Therefore, the force which is responsible for the deviation $HN$ (see Fig. \ref{fig : spin_distributions}) of the particle from the straight line is proportional to the magnetic moment $M_z$ and is collinear to axis $Oz$.\\

\noindent As the magnetic moments of the various particles are uniformly distributed between $+|M|$ and $-|M|$ it is equally possible to find particles having magnetic moment between $+|M|$ and $-|M|$. Thus one expects that the particles beam will generate on plate $P$ an equiprobable symmetric distribution diagram as shown with dashed line in Fig. \ref{fig : spin_distributions}.  However, in reality one observes two different distributions centered at points $N_1$ and $N_2$, which means that the particle's magnetic moment $M_z$ along axis z can take only two distinct values (magnetic moment eigenvalues or spin values).

\noindent The results of the Stern-Gerlach experiment lead to the following conclusion: if some one measures the component of the kinetic moment $\Gamma_z$ of the particle (which is proportional to the magnetic moment $M_z$) then only one of the values corresponding to deviations $HN_1$ and $HN_2$ can be found. Therefore, the classical description of a magnetic moment vector $M$ the measurements of which can take any value has to be abandoned. Consequently, the magnetic moment $M_z$ is a variable, with spectrum that can take only two eigenvalues (e.g. $\pm {1 \over 2}$).\\

\subsection{Measurement operators in the spin state-space\\}

\noindent It has been shown that the eigenvalues of the particle's magnetic moment are ${\pm}{1 \over 2}$ or ${\pm}{\hbar}{1 \over 2}$. The corresponding eigenvectors are denoted as $|+>$ and $|->$. Then the relation between eigenvectors and eigenvalues is given by ${S_z}|+>=+({\hbar/2})|+>$, ${S_z}|->=+({\hbar/2})|->$, which means that the measurement of the Stern-Gerlach experiment shows only the two possible eigenvalues of the magnetic moment. In general the particle's state, with reference to the spin eigenvectors, is described by

\begin{equation} \label{entangled_states}
\begin{tabular}{c}
$|\psi>={\alpha}|+>+{\beta}|->$ \\
$ $\\
$\textit{with} \ \ \ |\alpha|^2+|\beta|^2=1$
\end{tabular}
\end{equation}

\noindent while matrix $S_z$ has the eigenvectors $|+>=[1,0]$ and $|->=[0,1]$ and is given by

\begin{equation} \label{spin_measurement_matrix_z}
S_z={{\hbar} \over 2}
\begin{pmatrix}
1 & 0 \\
0 & -1
\end{pmatrix}
\end{equation}

\noindent Similarly, if one assumes components of magnetic moment along axes $x$ and $z$, one obtains the other two measurement (Pauli) operators

\begin{equation} \label{spin_measurement_matrix_x}
S_x={{\hbar} \over 2}
\begin{pmatrix}
0 & 1 \\
1 & 0
\end{pmatrix}, \ \
S_y={{\hbar} \over 2}
\begin{pmatrix}
0 & -i \\
i & 0
\end{pmatrix}
\end{equation}

\noindent When the eigenvalue of the magnetic moment of the particle $M_z$ is at state $|+>$ or at state $|->$ the particle follows a well-defined trajectory as shown in Fig. \ref{fig : spin_distributions}. When the spin's state is a linear superposition of $|+>$ and $|->$ it is no longer possible to predict the trajectory of the particle.  Therefore, there is a certain probability amplitude that the particle is found in one out of two trajectories. The probability to locate the particle at a specific point of surface $P$ (particle's wave-function) has non-zero values at two different areas which are designated by points $N_1$ and $N_2$. Therefore, the particle may either appear round point $N_1$ or round point $N_2$. However it is not possible to predict the particle's position with certainty, although the initial conditions are known (bifurcation). It is noted that in Fig. \ref{fig : spin_distributions} there are not two different particles, but one particle, with a probability wave-function that consists of two parts centered at points $N_1$ and $N_2$. \\

\subsection{The spin eigenstates define a two-level quantum system\\}

\noindent The spin eigenstates correspond to two different energy  levels. A neutral particle is considered in a magnetic field of intensity $B_z$. The particle's magnetic moment  $M$ and the associated kinetic moment $\Gamma$ are collinear and are related to each-other through the relation $M={\gamma}{\Gamma}$. The potential energy of the particle is $W=-{M_z}{B_z}=-{\gamma}{B_z}{\Gamma_z}$. Variable $\omega_0=-{\gamma}{B_z}$ is introduced, while parameter $\Gamma_z$ is substituted by the spin's measurement operator $S_z$.\\

\noindent Thus the Hamiltonian $H$ which describes the evolution of the spin of the particle due to field $B_z$ becomes $H_0={\omega_0}{S_z}$, and the following relations between eigenvectors and eigenvalues are introduced:

\begin{equation}
\begin{tabular}{c}
$H|+>=+{{\hbar}{\omega_0} \over 2}|+>, \ \ H|->=+{{\hbar}{\omega_0} \over 2}|->$
\end{tabular}
\end{equation}

\noindent Therefore, one can distinguish 2 different energy levels (states of the quantum system)

\begin{equation} \label{energy_levels}
\begin{tabular}{c}
$E_{+}=+{{{\hbar}{\omega_0}} \over 2},  \ \ E_{-}=-{{{\hbar}{\omega_0}} \over 2}$
\end{tabular}
\end{equation}

\noindent By applying an external magnetic field the probability of finding the particle's magnetic moment at one of the two eigenstates (spin up or down) can be changed. This can be observed for instance in the Nuclear Magnetic Resonance (NMR) model [\ref{TanDiuLal98}].\\

\section{Stochastic mechanics formulation of Schr\"{o}dinger's equation\\} \label{section :
Equivalence_Schrodinger_diffusion}

\subsection{Particles' motion can be written as a diffusion process\\} \label{subsection : formulation_Schrodinger_diffusion}

\noindent  Schr\"{o}dinger's equation, which was described
in Eq. (\ref{Schrodingers_equation}), can be transformed into a diffusion equation by substituting
variable $it$ with $t$ [\ref{Far06}]. This change of variable results in the diffusion
equation

\begin{equation} \label{diffusion_equivalent_Schrodinger}
{{{\partial}\rho} \over {{\partial}t}}=[{1 \over
2}{\sigma^2}{{\partial^2} \over {{\partial}x^2}}-V(x)]\rho
\end{equation}

\noindent Eq. (\ref{diffusion_equivalent_Schrodinger}) can be also
written as ${{{\partial}\rho} \over {{\partial}t}}=-H\rho$, where
$H$ is the associated Hamiltonian and the solution is of the form
${\rho}(x,t)=e^{-tH}{\rho}(x)$, and variable $\sigma^2$ is a
diffusion constant. The probability density function $\rho$
satisfies also the \textit{Fokker-Planck} partial differential
equation [\ref{Far06}]

\begin{equation} \label{Fokker_Planck}
{{{\partial}{\rho}} \over {{\partial}t}}=[{1 \over
2}{\sigma^2}{{{\partial}^2 \over {\partial}x^2}}-{{\partial} \over
{{\partial}x}}u(x)]{\rho}
\end{equation}

\noindent where $u(x)$ is the \textit{drift function}, i.e. a
function related through derivative to the external potential $V$.

\noindent Now, as known from quantum mechanics, particle's probability
density function $\rho(x)$ is a wave-function for which holds
$\rho(x)=|\psi(x)|^2$ with $\psi(x)={\sum_{i=0}^N}{c_k}\psi_k(x)$,
where $\psi_k(x)$ are the associated eigenfunctions
[\ref{TanDiuLal98}]. It can be assumed that
$\rho_0(x)=|{\psi_0^2}(x)|$, i.e. the p.d.f includes only the
basic mode, while higher order modes are truncated, and the drift
function $u(x)$ of Eq. (\ref{Fokker_Planck}) is taken to be
$u(x)={1 \over 2}{\sigma^2}{1 \over
{\rho_0(x)}}{{\partial}{{\rho_0}(x)} \over {{\partial}x}}$ [\ref{Far06}].
Thus, it is considered that the initial probability
density function is $\rho(x)={\rho_0}(x)$, which is independent of
time. This means that the p.d.f. remains
independent of time and the examined diffusion process is a
stationary one, i.e. $\rho(x,t)=\rho_0(x) \ \forall t$. A form of
the probability density function for the stationary diffusion is
that of shifted, partially overlapping Gaussians [\ref{Far06}].\\

\noindent Continuing from Fokker-Planck's equation, given in Eq. (\ref{Fokker_Planck}),
the \textit{Ornstein-Uhlenbeck} diffusion
is obtained which is a model of the Brownian motion [\ref{BaN93}]. The particle tries to return to
the equilibrium $x=0$ under the influence of a linear force, i.e.
there is a spring force applied to the particle as a result of the
potential $V(x)$. The corresponding phenomenon in quantum
mechanics is that of the quantum harmonic oscillator (Q.H.O.)
[\ref{TanDiuLal98}].  Assuming a
stationary p.d.f., i.e.
$\rho(x)=\psi_0(x)^2={C^2}e^{-{{\omega}{x^2} \over {2\sigma^2}}}$,
the force applied to the particle due to the harmonic potential (drift)
$V(x)$ is found to be $u(x)={\sigma^2}{1 \over {\psi_0(x)}}{{{\partial}{\psi_0}(x)}
\over {{\partial}x}} \Rightarrow u(x)=-{\omega}x$, which means that the drift is a spring force applied to
the particle and which aims at leading it to an equilibrium
position.\\

\noindent Now, a kinematic model for the particles will be derived, in the form of
\textit{Langevin's} equation. The stochastic differential equation for the position of the particle is [\ref{Far06}]:

\begin{equation} \label{Langevin_equation_position}
dx(t)=u(x(t))dt+dw(t)
\end{equation}

\noindent where $u(x)=-kx$  is the drift function, and is a spring force generated
by the harmonic potential $V(x)=kx^2$, which
tries to bring the particle to the equilibrium $x=0$. The term $w(t)$
denotes a random force (due to interaction with other particles) and results in
a Wiener walk. Knowing that the Q.H.O. model imposes to the particle
the spring force $u(x)=-{\omega}x$, Langevin's equation described in Eq.
(\ref{Langevin_equation_position}), becomes

\begin{equation} \label{Langevin_equation}
dx(t)=-{\omega}x(t)dt+dw(t)
\end{equation}

\noindent  Eq. (\ref{Langevin_equation}) is a
generalization of gradient algorithms based on the ordinary
differential equation (O.D.E) concept, where the gradient
algorithms are described as trajectories towards the equilibrium
of an ordinary differential equation
[\ref{BenMetPri90}].\\

\subsection{Particle's spin in stochastic mechanics\\} \label{subsection : particle_spin_stochastic_mechanics}

\noindent For a particle described in classical quantum mechanics by Schr\"{o}dinger's equation, the spin represented the eigenvalues of the particle's magnetic moment (projection on the $z$ axis) and the associated eigenstates were
denoted as $|+>$ (spin up) and $|->$ (spin down). A representation for spin can be also obtained in terms of stochastic mechanics. The particle is described not only in the $R^3$ space by its cartesian coordinates, but is also described in the $SU(2)$ space, i.e. it is described also by the spin variables which take two discrete values.\\

\noindent As explained in the description of the Stern-Gerlach experiment in subsection \ref{subsection : particle_motion_magnetic_field}, particles with spin $\pm{1 \over 2}$ are emitted from a source and pass through a magnetic field $B_z$. Then, the kinetic moment of the particle is proportional to its magnetic moment, and due to its interaction with the magnetic field generates a force which causes scattering of the particle's trajectory $F={\nabla}({M}{B_z})$. If the particle's path deviates towards the positive (negative) direction of the gradient then one can conclude that the particle's spin is at state $|+>$, or $|+>$. The particle's trajectory shows the discrete spin eigenvalues, while the particle's speed shows the discrete evergy levels $E_{+}$ and $E_{-}$ which correspond to the spin eigenvalues. It has been shown that starting from Schr\"{o}dinger's equation and continuing to Focker-Planck and Ornstein-Uhlenbeck equations one obtains Langevin's stochastic differentinal equation, i.e. $d{x_t}=b(x,t){dt}+d{w_t}$. It has been also shown that for long time $t$, $b(x,t){\simeq}{x \over t}$, i.e. in $b(x,t)=k{\cdot}x$ one can consider $k={1 \over t}$, thus ones obtains the SDE [\ref{Far06}]

\begin{equation} \label{modified_Langevin_SDE}
d{x_t}={1 \over t}{x_t}{dt}+d{w_t}
\end{equation}

\noindent and based on the so-called "martingale convergence theorem" it has been proven that the limit ${\text{lim}_{t{\rightarrow}\infty}}{x_t \over t}$ exists and is $p_{+}={\text{lim}_{t{\rightarrow}\infty}}{x_t \over t}$.
This limit is the kinetic moment which corresponds to the magnetic moment eigenstate $|+>$ and to the magnetic moment eigenvalue 'spin-up'. Moreover, it has been shown that for every measurable subset $A{\in}{R^3}$, the probability $P_{+}(p_t \in A)$ is equal to the quantum mechanical probability that the final moment of the particle belongs in $A$. Thus, in stochastic mechanics a way to measure the moment is through the limit $p_{+}={\text{lim}_{t{\rightarrow}\infty}}{x_t \over t}$.\\

\noindent Consequently, in stochastic mechanics, the different energy eigenstates of the particle can be conceived according to the Stern-Gerlach experiment as follows: the particle has initial moment $p_{-}$, and while it approaches to the surface $P$ in which points $N_1$ and $N_2$ belong (see Fig. \ref{fig : spin_distributions}), the particle's trajectory becomes straight and the final moment becomes $p_{+}$. Then, the only possible values for the change of energy are

\begin{equation} \label{change_energy_eigenstates}
\begin{tabular}{c}
$m(|p_{+}|^2-|p_{-}|^2), \ \text{i.e}\ \  E_{+}-E_{-}$\\
\end{tabular}
\end{equation}

\noindent where, $E_j$ is the Hamiltonian's eigenvalue, as explained in Eq. \ref{energy_levels}. Therefore, in stochastic mechanics the stage of the particle's motion at which its trajectory becomes straight and its moment stabilizes at the final value $p_{\pm}$ can be considered as a collapse of the particle's wave function.\\

\section{Open-loop control scheme for a multi-particle system\\}

\noindent As explained in subsection \ref{subsection : particle_motion_magnetic_field} the change in the particle's moment means transition between the particle's energy eigenstates. This can be a manner to control the probability to find the particle's magnetic moment in one of the spin's eigenstates. The particle's kinematic model has been described by Eq. (\ref{Langevin_equation}), and here it will be formulated as follows: the motion of the particle is described by the stochastic oscillator model $m{{{d^2}x} \over {dt^2}}+c{{dx} \over {dt}}=f$, where $x$ is the position coordinate, $m$ is the mass, $c$ is the coefficient of viscous friction, and $f$ is the aggregate force acting on the particle [\ref{Ast06}].  Defining  $x_1=x$ and $x_2={{dx} \over {dt}}$, particle's motion can be written in state-space form ${{d{x_1}} \over {dt}}=x_2$, and ${{d{x_2}} \over {dt}}=-{c \over m}{x_2}+{1 \over m}f$. Keeping the second of the state-space equations one has the Langevin equation for the particle's velocity

\begin{equation} \label{Langevin_velocity}
\begin{tabular}{c}
${{d{v}} \over {dt}}=-{c \over m}{v}+{1 \over m}f$.
\end{tabular}
\end{equation}

\noindent Next, an open-loop control, based on \textit{flatness-based control} theory, will be  applied to the particles: \\

\noindent \textit{Definition}: The system $\dot{x}=f(x,u), \ x \in \ R^n, u \in \ R^m$
is differentially flat if there exist relations $h: \ {R^n}{\times}({R^m})^{r+1}{\rightarrow}{R^m}$,
$\phi: \ {(R^m)^r}{\rightarrow}{R^n}$  and $\psi: \ (R^m)^{r+1} \rightarrow {R^m}$, such that
$y=h(x,u,\dot{u},\cdots,u^{(r)})$, $x=\phi(y,\dot{y},\cdots,y^{(r-1)})$ and $u=\psi(y,\dot{y},\cdots,y^{(r-1)},y^{(r)})$.
This means that all system dynamics can be expressed as a function of the flat output and its derivatives, therefore the
state vector $x$ and the control input $u$ can be written as $x(t)=\phi(y(t),\dot{y}(t),\cdots,y^{(r)}(t))$ and
$u(t)=\psi(y(t),\dot{y}(t),\cdots,y^{(r+1)}(t))$ [\ref{FliMou99}],[\ref{Rou05}],[\ref{MarRou99}].

\noindent If a control term $u^i$ is introduced in Eq. (\ref{Langevin_velocity}) this can be written as:

\begin{equation} \label{flat_ith_robot}
\begin{tabular}{c}
$\dot{v}^i=-\omega{v}^i+u^i+\eta^i$
\end{tabular}
\end{equation}

\noindent where $-\omega{x}^i$ is the drift term due to an external potential, $u^i$ is the external control and $\eta^i$ is a disturbance term due to interaction with the rest $N$-1 particles, or due to the existence of noise. Then it can be easily shown that the system of Eq. (\ref{flat_ith_robot}) is differentially flat, while an appropriate flat output can be chosen to be $y=v^i$. Indeed all system variables, i.e. the elements of the state vector and the control input can be written as functions of the flat output $y$, and thus the model that describes the $i$-th particle is differentially flat.\\

\noindent A control input that makes the $i$-th particle track the reference trajectory $y_r^i$ is given by

\begin{equation} \label{flatness_control_ith_robot}
\begin{tabular}{c}
$u^i={\omega}{v^i_r}+{\dot{v}^i_r}+u_c^i$,
\end{tabular}
\end{equation}

\noindent where $v_r^i$ is the reference velocity profile for the $i$-th particle, and $\dot{v}_i^r$ is the derivative of the $i$-th desirable velocity. Moreover $u_c^i=-\eta^{i}$ stands for an additional control term which compensates for the effect of the noise $\eta_i$ on the $i$-th particle. Thus, if the disturbance $\eta_i$ that affects the $i$th-particle is adequately approximated it suffices to set $u_c^i=-\eta_i$. The application of the control law of Eq. (\ref{flatness_control_ith_robot}) to the model of Eq. (\ref{flat_ith_robot}) results in the error dynamics $\dot{v}^i=\dot{v}^i_r-{\omega}{v^i}+{\omega}{v^i_r}+\eta^i-u_c^i \Rightarrow$, i.e. $\dot{v}^i-\dot{v}^i_r+{\omega}(v_i-v_i^r)=\eta^i+u_c \Rightarrow \dot{e}^i+{\omega}e^i=\eta_i+u_c$. Thus, if $u_c=-\eta_i$ then $\text{lim}_{t{\rightarrow}\infty}e=0$.\\

\noindent Next, the case of the $N$ interacting particles will be examined. The control law that makes the mean of the
multi-particle system follow a desirable velocity profile $E\{v_r^i\}$ can be derived. The kinematic model of the mean of the multi-particle system is given by

\begin{equation} \label{kinematic_model_mean}
\begin{tabular}{c}
$E\{\dot{v}^i\}=-{\omega}E\{v^i\}+E\{u^i\}+E\{\eta^i\}$
\end{tabular}
\end{equation}

\noindent $i=1,\cdots,N$, where $E\{v^i\}$ is the mean value of the particles velocity, $E\{\dot{v}^i\}$ is the mean acceleration, $E\{\eta^i\}$ is the average of the disturbance signal and $E\{u^i\}$ is the mean control input. The open-loop controller is selected as:

\begin{equation} \label{control_input_mean}
E\{u^i\}={\omega}E\{v^i\}_r+E\{\dot{v^i}\}_r-E\{\eta^i\}
\end{equation}

\noindent where $E\{v^i\}_r$ is the desirable mean velocity. Assuming that for the mean of the particles system holds
$E\{\eta^i\}=0$, then the control law of Eq. (\ref{control_input_mean}) results in the error dynamics $E\{\dot{e}^i\}+{\omega}E\{e^i\}=0$, which assures that the mean particles' velocity will track the desirable velocity profile, i.e. $\text{lim}_{t{\rightarrow}\infty}{E\{e^i\}=0}$.\\

\section{Conclusions\\}

\noindent The paper has proposed the spin-orbit interaction as a model of quantum computation. The stochastic mechanics formulation of Schr\"{o}dinger's equation was introduced. This enables to consider the control of the entangled spin state as a problem of motion control of the mean of a particles ensemble along a reference trajectory. It is shown that using an appropriate open-loop control (which can be created by a magnetic field) the mean velocity of the particles can track a desirable velocity profile. This in turn implies that the quantum state represented by the mean particles moment can be also controlled.\\

\end{document}